\documentclass[twocolumn,prd,preprintnumbers,amsmath,amssymb]{revtex4}

\usepackage{graphicx}
\usepackage{dcolumn}
\usepackage{bm}
\usepackage{epsfig}
\usepackage{amssymb}
\usepackage{amsmath}

 \def\tskip{\setlength{\tskip}{5pt}}
\def\colwidth{\setlength{\colwidth}{3.5in}}

\newcommand{\lsim}{\mathrel{\hbox{\rlap{\lower.55ex\hbox{$\sim$}} \kern-.3em \raise.4ex \hbox{$<$}}}}
\newcommand{\gsim}{\mathrel{\hbox{\rlap{\lower.55ex\hbox{$\sim$}} \kern-.3em \raise.4ex \hbox{$>$}}}}
\newcommand{\beq}{\begin{equation}}
\newcommand{\eeq}{\end{equation}}
\newcommand{\beqa}{\begin{eqnarray}}
\newcommand{\eeqa}{\end{eqnarray}}

\begin{document}

\title{Detecting Relic Gravitational Waves in the CMB: Comparison of Planck and Ground-based Experiments}

\author{Wen Zhao} \email{Wen.Zhao@astro.cf.ac.uk}
\affiliation{School of Physics and Astronomy, Cardiff University,
Cardiff, CF24 3AA, United Kingdom \\
Wales Institute of Mathematical and Computational Sciences,
Swansea, SA2 8PP, United Kingdom \\ Department of Physics,
Zhejiang University of Technology, Hangzhou, 310014, People's
Republic of China }
\author{Wei Zhang}
\affiliation{National Astronomical Observatories, Chinese Academy of Sciences, Beijing, 100012, People's Republic of China}

\date{\today}

\begin{abstract}
We compare the detection abilities for the relic gravitational
waves by two kinds of forthcoming cosmic microwave background
radiation (CMB) experiments, space-based Planck satellite and the
various ground-based experiments. Comparing with the ground-based
experiments, Planck satellite can observe all the CMB power
spectra in all the multipole range, but having much larger
instrumental noises. We find that, for the uncertainty of the
tensor-to-scalar ratio $\Delta r$, PolarBear (II) as a typical
ground-based experiment can give much smaller value than Planck
satellite. However, for the uncertainty of the spectral index
$\Delta n_t$, Planck can give the similar result with PolarBear
(II). If combining these two experiments, the value of $\Delta
n_t$ can be reduced by a factor $2$. For the model with $r=0.1$,
the constraint $\Delta n_t=0.10$ is expected to be achieved, which
provides an excellent opportunity to study the physics in the
very early universe. We also find the observation in the largest
scale ($\ell<20$) is very important for constraining the spectral
index $n_t$. So it is necessary to combine the observations of the
future space-based and ground-based CMB experiments to determine
the relic gravitational waves.
\end{abstract}


\maketitle


\section{Introduction \label{s-introduction}}

Relic gravitational waves (RGWs) are generated in the very early
Universe due to the superadiabatic amplification of zero point
quantum fluctuations of the gravitational field \cite{a1,s2}, and
freely evolve in the whole stage of the Universe
\cite{zhao,others}. So RGWs carry invaluable information about
early history of our Universe inaccessible to any other medium.

Detection of RGWs is rightly considered a highest priority for the
upcoming cosmic microwave background radiation (CMB) experiments.
The current CMB experiments are yet to detect a definite signature
of RGWs \cite{experiment1}. In the near future, the successive
generation of the experiments, including the space-based Planck
satellite \cite{Planck} and the various ground-based experiments
\cite{bicep,polarbear,quiet,experiment2} together with a host of
balloon-borne experiments \cite{balloon} will provide an
increasing sensitive measurement of RGWs.

The space-based experiments, as COBE, WMAP and Planck satellites,
can remove atmospheric noises and observe the fairly cleaned CMB
temperature and polarization anisotropy fields. In addition, the
space-based experiments provide the unique opportunity to detect
the CMB power spectra in the largest scale ($\ell<20$) by
surveying the full sky.

At the same time, the CMB polarization field can also be observed
by the ground-based experiments. Since the atmospheric emission is
not expected to be linearly polarized \cite{keating1998},  by
integrating deeply on the relatively small patches of sky, it is
possible to make a measurement of the polarization anisotropies
with a comparable signal-to-noise ratio to a satellite experiment
on all but the largest angular scales.

In this letter, we shall investigate the detection abilities for
the RGWs, by the observations of the forthcoming generation of the
ground-based and space-based experiments. By calculating the
constraints on  the tensor-to-scalar ratio $r$ and the tensor
spectral index $n_t$, we shall compare the detection abilities of
these two types of experiments. We also investigate the potential
improvement of the detection ability by combining the observations
of them.


\section{Review of primordial perturbations, CMB and noises \label{s-spectra}}

\subsection{Primordial perturbation power spectra}
The main contribution to the temperature and polarization
anisotropies of the CMB comes from two types of cosmological
perturbations, density perturbations (also known as the scalar
perturbations) and RGWs (also known as the tensor perturbations).
The primordial power spectra of these perturbations are usually
assumed to be power laws, which is a generic prediction of a wide
range of scenarios of the early Universe \cite{lyth}. If we ignore
the running of the spectral indices, the primordial spectra can be
written as the following simple forms
\begin{equation}
P_{{s}}(k)=A_s(k_0)\left({k}/{k_0}\right)^{n_s-1},
\end{equation}
\begin{equation}\label{pt}
P_{t}(k)=A_t(k_0)\left({k}/{k_0}\right)^{n_t},
\end{equation}
where $k_0$ is the pivot wavenumber, which can be arbitrarily chosen. $n_s$ and $n_t$ are the scalar and tensor spectral indices.
The tensor-to-scalar ratio is defined by
\begin{equation}
r(k_0)\equiv \frac{A_t(k_0)}{A_s(k_0)}.
\end{equation}
For a fixed $A_{s}$, the primordial power spectra of RGWs are
completely determined by two parameters $r$ and $n_t$, if a
power-law form in (\ref{pt}) is assumed. The simplest single-field
slow-roll inflationary models predict a consistency relation
between $r$ and $n_t$ \cite{lyth}: $n_t=-r/8$. However, this
consistency relation is incorrect for other inflationary models
\cite{othermodels}. So the determination the parameters $r$ and
$n_t$ by the observations, provides an excellent opportunity to
distinguish various inflationary-type models.

Since in this letter, we are primarily interested in the
parameters of the RGW field, in the analysis below we shall work
with a fixed cosmological background model. More specifically, we
shall work in the framework of $\Lambda$CDM model, and keep the
background cosmological parameters fixed at the values determined
by a typical model \cite{typical}
\begin{eqnarray}
\label{background}
\begin{array}{c}
h=0.732,~\Omega_b
h^2=0.02229,\\
\Omega_{m}h^2=0.1277,~\Omega_{k}=0,~\tau_{reion}=0.089.
\end{array}
\end{eqnarray}
Furthermore, in order to show the results in the figures, we adopt the following parameters of the density perturbations and tensor spectral index,
\begin{eqnarray}
\label{dp}
A_s=2.3\times
10^{-9},~n_s=1,~n_t=0.
\end{eqnarray}

\subsection{CMB power spectra and their estimators}

Let us turn our attention to the CMB field. Density perturbations
and gravitational waves produce temperature and polarization
anisotropies in the CMB characterized by four angular power
spectra $C_{\ell}^{T}$, $C_{\ell}^{E}$, $C_{\ell}^{B}$ and
$C_{\ell}^{C}$ as functions of the multipole number $\ell$
\cite{a8,grishchuk-cmb,a10,a11}. Here $C_{\ell}^{T}$ is the power
spectrum of the temperature anisotropies, $C_{\ell}^{E}$ and
$C_{\ell}^{B}$ are the power spectra of the so-called $E$-mode and
$B$-mode polarizations and $C_{\ell}^{C}$  is the power spectrum of
the temperature-polarization cross correlation.

In the linear theory, the various power spectra $C_{\ell}^{Y}$
(where $Y=T,E,B$ or $C$) can be presented in the following form
\begin{eqnarray}\label{c-sum}
C_{\ell}^{Y}=C_{\ell,s}^{Y}+C_{\ell,t}^{Y},
\end{eqnarray}
where $C_{\ell,s}^{Y}$ are the power spectra due to the density
perturbations, and $C_{\ell,t}^{Y}$ are the power spectra due to
RGWs.

The CMB power spectra $C_{\ell}^{Y}$ are theoretical constructions
determined by ensemble averages over all possible realizations of
the underlying random process. However, in real CMB observations,
we only have access to a single sky, and hence to a single
realization. In order to obtain information on the power spectra
from a single realization, it is required to construct estimators
of power spectra $D_{\ell}^Y$ \cite{ours1}. The probability
distribution functions for the estimators are detailed described
in \cite{ours1}, which predicts the expectation values of the
estimators \begin{equation} \langle D_{\ell}^Y\rangle=C_{\ell}^Y,
\end{equation}
and the standard deviations \cite{ours1}
\begin{eqnarray}\label{variance}
\begin{array}{c}
(\sigma_{D_{\ell}^X})^2=\frac{2(C_{\ell}^X+N_{\ell}^X)^2}{(2\ell+1)f_{\rm sky}},~~(X=T,E,B)\\
(\sigma_{D_{\ell}^C})^2=\frac{(C_{\ell}^T+N_{\ell}^T)(C_{\ell}^E+N_{\ell}^E)+(C_{\ell}^C)^2}{(2\ell+1)f_{\rm sky}},
\end{array}
\end{eqnarray}
where $f_{\rm sky}$ is the cut-sky factor, and $N_{\ell}^X$ are
the noise power spectra, determined by the specific experiments
\footnote{ For the actual observations of the ground-based
experiments, we always have to bin the observed data to keep the
data at different $\ell$ being uncorrelated \cite{hobson}.
However, the bin does not affect our conclusion in this paper. }.

\subsection{Noise power spectra}
Considering an experiment with multiple frequency channels, the
total instrumental noise power spectra can be approximately
presented as (see for instance \cite{ground}),
\begin{equation}\label{instrumental}
N_{\ell}^{X}=\left(\sum_{c}\frac{1}{N_{\ell,c}^{X}}\right)^{-1}.
\end{equation}
Here $N_{\ell,c}^{X}$ is the noise power spectrum for the
individual frequency channel, which is given by
\begin{equation}
N_{\ell,c}^{X}=(\sigma_{pix}^X\cdot\theta_{\rm
F})^2\cdot\exp\left[\ell(\ell+1)\frac{\theta_{\rm F}^{~2}}{8\ln
2}\right].
\end{equation}
In this formula, $\theta_{\rm F}$ is the full width half maximum
(FWHM) beam size. The pixel noises $\sigma_{pix}^X$ depend on the
survey design and the instrumental parameters.

\begin{figure}[t]
\centerline{\includegraphics[width=8cm]{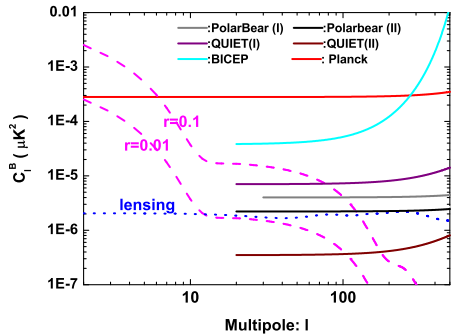}} \caption{This
figure shows the instrumental noise power spectrum $N_{\ell}^B$
for Planck satellite and various ground-based experiments (solid
lines). For the comparison, we also plot the power spectra
$C_{\ell}^{B}$ in the models with $r=0.1$ and $r=0.01$ (magenta
dashed lines). The blue dotted line denotes the power spectra
$C_{\ell}^{B}$ generated by cosmic lensing. }\label{figure0}
\end{figure}

First, let us focus on the space-based Planck satellite. In this
paper, we consider four frequency channels for the Planck
satellite, which are listed in Table \ref{table1}. After two full
sky survey (14 months), the pix noises $\sigma_{pix}^X$ are
expected to be the values listed in Table \ref{table1} for
different channels.  In FIG.\ref{figure0}, we plot the
instrumental noise power spectrum $N_{\ell}^B$, as a comparison
with the values of CMB power spectrum $C_{\ell}^B$ in the model
with $r=0.1$ and $r=0.01$. We find, in the model of $r=0.1$, the
value of $C_{\ell}^B$ is larger than that of $N_{\ell}^{B}$ only
at the largest scale. So Planck satellite can detect RGWs mainly
by the observation in this largest scale, which will be clearly
shown in the following section.

For a ground-based experiment with $N_{d}$ detectors, a solid
angle per pixel $\theta_{\rm F}^{~2}$ and a sensitivity NET, we
assume it will survey an area $4\pi f_{\rm sky}$ in the
integration time $t_{obs}$. The pixel polarization noises are
\begin{equation}
(\sigma_{pix}^E)^2=(\sigma_{pix}^B)^2=\frac{(\sqrt{2}{\rm
NET})^2\cdot 4\pi f_{\rm sky}}{t_{obs}N_{d}~\theta_{\rm F}^{~2}}.
\end{equation}
In this paper, we shall discuss five kinds of ground-based
experiments: BICEP, PolarBear (I) and (II), QUIET (I) and (II).
The instrumental parameters for these experiments are given in the
Appendix \ref{appendix}. Notice that, the ground-based experiments
are only sensitive to the polarizations. Since the ground-based
experiments can only survey a small part of the full sky, it
cannot encode the information of CMB field in the very large
scale. From FIG.\ref{figure0}, we find that ground-based
experiments have much smaller noise power spectra than Planck
satellite.

We should notice that, cosmic lensing can convert the $E$-mode
polarization into $B$-mode (see \cite{lens} for a review). So the
$B$-mode spectrum due to RGWs will be contaminated by a cosmic
lensing contribution. The lensed $C_{\ell}^{B}$ is also shown in
FIG.\ref{figure0}, which can be treated as a part of the total
noise power spectrum $N_{\ell}^{B}$ as well as instrumental noise
power spectra in (\ref{instrumental}).

In addition to the instrumental noises and lensing noise, various
foregrounds, such as the synchrotron and dust, are also the
important contaminations in the CMB observation. It is hoped that
the multifrequency observations and the hard work by astronomical
community might allow future experiments can reduce the foreground
noises in a very accurate level (see for instance
\cite{foreground}). So in the following discussion, we shall not
consider this kind of contamination.

\begin{table}
\caption{Instrumental parameters for Planck satellite
\cite{Planck}. }
\begin{center}
\label{table1}
\begin{tabular}{|c|c|c|c|}
  \hline
   Band center~[GHz] & 100& 143 & 217 \\
  \hline
  $\sigma_{pix}^T$~[$\mu$K]  & 6.8 &   6.0 & 13.1  \\
  \hline
   $\sigma_{pix}^E$ and $\sigma_{pix}^B$~[$\mu$K] &10.9 &  11.5 &26.8 \\
    \hline
   FWHM~[arcmin] & $9.5$ &  $7.1$ & $5.0$ \\
  \hline
  \multicolumn{1}{|c|}{$f_{\rm sky}$} & \multicolumn{3}{c|}{$0.65$}\\
           \hline
   \multicolumn{1}{|c|}{$\ell$ range} & \multicolumn{3}{c|}{$2\sim1000$}\\
             \hline
  \multicolumn{1}{|c|}{Integration time} & \multicolumn{3}{c|}{$14$~Months}\\
  \hline
\end{tabular}
\end{center}
\end{table}


\section{Determination of RGWs by CMB observations\label{s3}}

In the previous works \cite{ours1,ours2,ours3}, we have discussed
how to best constrain the parameters of the RGWs, i.e. $r$ and
$n_t$, by the CMB observation. In the paper \cite{ours3}, we found
that, in general the constraints on $r$ and $n_t$ correlate with
each other. However, if we consider the tensor-to-scalar ratio at
the best-pivot wavenumber $k_t^*$, i.e. $r\equiv r(k_t^*)$, the
constraints on $r$ and $n_t$ becomes independent of each other,
and the uncertainties $\Delta{r}$ and $\Delta{n_t}$ have the
minimum values. In the work \cite{ours3}, we have derived the
formulas to calculate the quantities: the best-pivot wavenumber
$k_t^*$, and the uncertainties of the parameters $\Delta r$ and
$\Delta {n_t}$. This provides a simple and quick method to
investigate the ability of the CMB observations for the detect of
RGWs. In this section, we shall briefly introduce these results.

It is convenient to define two quantities as below,
\begin{equation}
a_{\ell}^Y\equiv\frac{C_{\ell,t}^{Y}}{\sigma_{D_{\ell}^Y}},~~b^*_{\ell}\equiv
\ln\left(\frac{\ell}{\ell_t^*}\right).
\end{equation}
Here $C_{\ell,t}^Y$ are the CMB power spectra generated by RGWs,
and $\sigma_{D_{\ell}^Y}$ are the standard deviations of the
estimators $D_{\ell}^Y$, which can be calculated by
Eq.(\ref{variance}). $\ell_t^*$ is the best-pivot multipole, which
is determined by solving the following equation \cite{ours3}:
\begin{equation}\label{formula1}
\sum_{\ell}\sum_{Y}a_{\ell}^{Y2}b_{\ell}^*=0.
\end{equation}
So the value of $\ell_t^*$ depends on the cosmological model, the
amplitude of RGWs, and the noise power spectra. The best-pivot
wavenumber $k_t^*$ relates to $\ell_t^*$ by the approximation
\cite{ours3},
\begin{equation}
k_t^*\simeq \ell_t^*\times 10^{-4}{\rm Mpc}^{-1}.
\end{equation}

In order to determine the constraints on the RGWs from the CMB
observation, we can consider two quantities, the signal-to-noise
ratio $S/N$ (which directly relate to $\Delta r$) and uncertainty
$\Delta n_t$. For a specific cosmological model and the noises,
these quantities can be calculated by the following formulas
\cite{ours1,ours2,ours3}
\begin{eqnarray}\label{formula2}
\begin{array}{c}
S/N\equiv {r}/{\Delta
r}=\sqrt{\sum_{\ell}\sum_{Y}a_{\ell}^{Y2}}, \\
\Delta n_t=1/\sqrt{\sum_{\ell}\sum_{Y}(a_{\ell}^Yb_{\ell}^*)^2}.
\end{array}
\end{eqnarray}

In this letter, in order to compare the detection abilities for
RGWs of Planck and the ground-based experiments, we shall consider
the following four cases:

{\bf Case A}: We only consider the observation of the
$B$-polarization by Planck satellite. So, in Eqs. (\ref{formula1})
and (\ref{formula2}), $Y=B$ and $\ell=2\sim1000$. The noise power
spectrum $N_{\ell}^B$ and cut sky factor $f_{\rm sky}$ are the
corresponding quantities for Planck satellite.

{\bf Case B}: We only consider the observation of the
$B$-polarization by the ground-based experiments.

{\bf Case C}: We consider the determination on the RGWs by
combining the $B$-polarization observations of Planck and the
various ground-based experiments. Since the power spectra in the
scale $\ell<20$ ($\ell<30$ for PolarBear (I)) can only be observed
by Planck satellite, we adopt the $N_{\ell}^B$ and $f_{\rm sky}$
as the the corresponding quantities for Planck satellite. In the
scale $\ell=20\sim1000$ ($\ell=30\sim1000$ for PolarBear (I)), we
only adopt the $N_{\ell}^B$ and $f_{\rm sky}$ as the corresponding
quantities for ground-based experiments.

{\bf Case D}: In addition to the $B$-polarization discussed in
{\bf Case C}, we also take into account the observations of the
other three power spectra, i.e. $Y=T,C,E$. For $Y=T$ and $C$, only
observed by Planck satellite, we consider the Planck noises and
cut sky factor.  For $Y=E$, similar with $Y=B$, Planck noise and
cut sky factor are considered for $\ell<20$ ($\ell<30$ for
PolarBear (I)), and the noises and cut sky factors of ground-based
experiments are considered for the other multipole scales.

\subsection{Best-pivot multipole $\ell_t^*$}

\begin{figure}[t]
\centerline{\includegraphics[width=8cm]{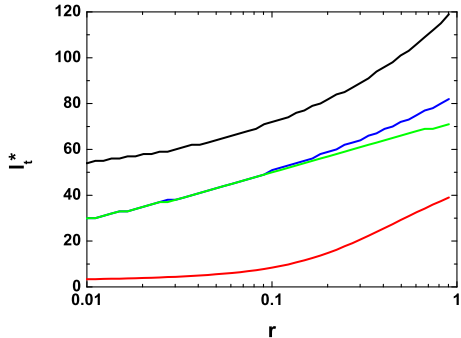}} \caption{This
figure shows the best-pivot multipole $\ell_t^*$ for Planck and
PolarBear (II) experiments in Case A (red line), Case B (black
line), Case C (blue line) and Case D (green line).}\label{figure1}
\end{figure}

\begin{figure}[t]
\centerline{\includegraphics[width=8cm]{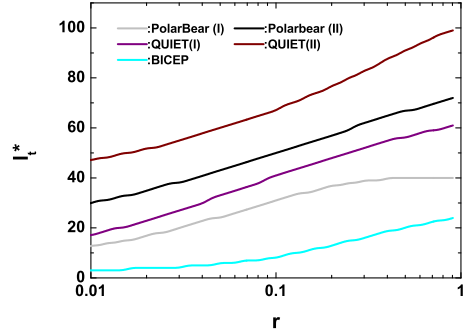}} \caption{This
figure shows the best-pivot multipole $\ell_t^*$ in Case D for the
various ground-based experiments.}\label{figurec1}
\end{figure}

First, let us discuss PolarBear (II) as a typical ground-based
experiment. By solving Eq.(\ref{formula1}),  in FIG.\ref{figure1}
we plot the best-pivot multipole $\ell_t^*$ as a function of
tensor-to-scalar ratio $r$ in the four cases.  In {\bf Case A},
the value of $\ell_t^*$ is always smaller than $40$, which is
because that, in {\bf Case A} the main contribution on the
constraint of RGWs only comes from the observation in the very
large scale, i.e. the reioniation peak of $B$-polarization power
spectrum \cite{ours1,ours2}.  In {\bf Case B}, $50<\ell_t^*<130$,
the best-pivot multipole is in the intermedial scale, which
reflects that PolarBear (II) constrains the RGWs mainly by the
observation in the intermedial scale.  As a combination of {\bf
Case A} and {\bf B}, in {\bf Case C} the value of $\ell_t^*$ is
focused on the range $40<\ell_t^*<90$. If we also consider the
other three power spectra, $T,E,C$ in {\bf Case D}, the value of
$\ell_t^*$ decreases a little when $r>0.1$, due to the
contribution of $T,E,C$ power spectra \cite{ours2}.

In FIG. \ref{figurec1}, we also plot the best-pivot multipole
$\ell_t^*$ as a function of $r$ in {\bf Case D}, where the various
ground-based experiments are considered. In all these cases, we
find that the value of $\ell_t^*$ increases with the increasing of
$r$. In general, the smaller instrumental noises follow the larger
$\ell_t^*$.

\subsection{Signal-to-noise ratio $S/N$}

\begin{figure}[t]
\centerline{\includegraphics[height=6cm,width=8cm]{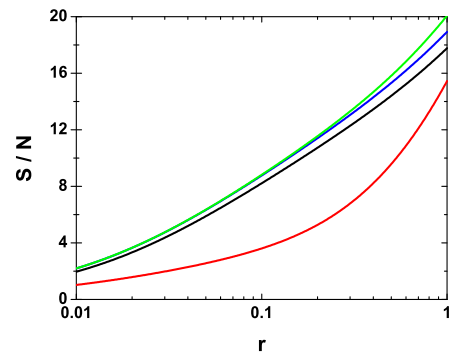}}
\caption{This figure shows the signal-to-noise ratio $S/N$ for
Planck and PolarBear (II) experiments in Case A (red line), Case B
(black line), Case C (blue line) and Case D (green
line).}\label{figure2}
\end{figure}

\begin{figure}[t]
\centerline{\includegraphics[height=6cm,width=8cm]{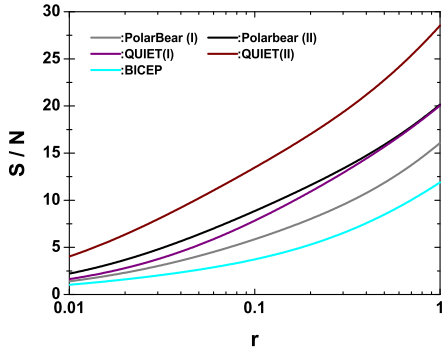}}
\caption{The figure shows the signal-to-noise ratio $S/N$  in Case
D for the various ground-based experiments.}\label{figurec2}
\end{figure}

FIG.\ref{figure2} presents the signal-to-noise ratio $S/N$ as a
function of $r$ for Planck and PolaeBear (II) experiments, which
are obtained by using the first formula in Eq.(\ref{formula2}). As
expected, in all these four cases, a larger $r$ predicts a larger
$S/N$. When $r=0.1$, $S/N=3.6$ for Planck satellite, and $S/N=8.4$
for PolarBear (II) experiment.  If we require that $S/N>3$,
$r>0.07$ must be satisfied for Planck satellite, and $r>0.02$ for
PolarBear (II) experiment. This figure shows that, PolarBear (II)
can give a much tighter constraint of $r$ than Planck satellite,
due to the much smaller noise level of the PolarBear (II)
experiment. Even if we combine Planck and PolarBear (II)
experiments, the constraint on $r$ cannot make obvious
improvement.

In FIG. \ref{figurec2}, we also plot the $S/N$ as a function of
$r$ in {\bf Case D}, where the various ground-based experiments
are considered.  As expected, the values $S/N$ strongly depend on
the instrumental noises and the sky survey factor $f_{\rm sky}$ of
the ground-based experiments. The QUIET (II) can very well detect
the signal of RGWs ($4$-$\sigma$ level), even for the model with
$r=0.01$,

\begin{figure}[t]
\centerline{\includegraphics[width=9cm]{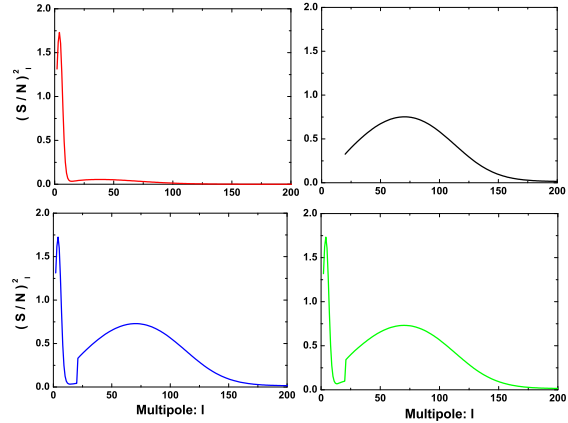}} \caption{In the
model of $r=0.1$ for Planck and PolarBear (II) experiment, we plot
the individual signal-to-noise ratio $(S/N)^2_{\ell}$ as a
function of multipole $\ell$ in Case A (red line), Case B (black
line), Case C (blue line) and Case D (green line).}\label{figure3}
\end{figure}

From the first formula in (\ref{formula2}), we find the total
signal-to-noise ratio can be written as
$(S/N)^2=\sum_{\ell}(S/N)_{\ell}^2$, where the individual
signal-to-noise ratio for the multipole $\ell$ is
$(S/N)_{\ell}^2=\sum_{Y}a_{\ell}^{Y2}$. FIG.\ref{figure3} presents
the quantity $(S/N)_{\ell}^2$ as a function of multipole $\ell$ in
the model with parameter $r=0.1$. This figure clearly shows that,
for the Planck satellite the constraint on $r$ mainly comes from
the observation in the reionization peak at $\ell<10$, and for
PolarBear (II) experiment, the constraint mainly comes from the
observation in the intermedial scale $20<\ell<150$. In {\bf Case
C} and {\bf D}, the function $(S/N)_{\ell}^2$ has two peaks, one
is at $\ell<10$, and the other is at $\ell\sim80$. Comparing with
the second peak, the first peak, due to the observation of Planck
satellite, is very narrow, and only contribute a fairly smalle
portion for the total $S/N$. From FIG.\ref{figure3}, we also find
the difference between {\bf Case C} and {\bf D} is only at the
range $10<\ell<20$, due to the observation of the $Y=T,E$ and $C$
power spectra.

\subsection{Uncertainty of spectral index $\Delta n_t$}

\begin{figure}[t]
\centerline{\includegraphics[width=8cm]{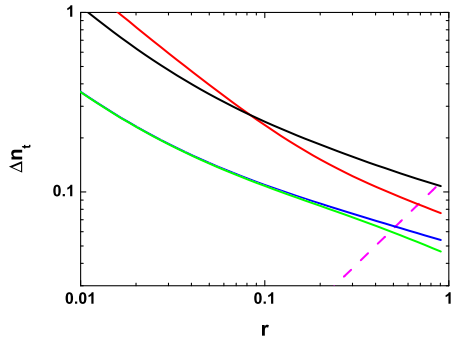}} \caption{This
figure shows the uncertainty $\Delta n_t$ for Planck and PolarBear
(II) experiments in Case A (red line), Case B (black line), Case C
(blue lines) and Case D (green lines). The dashed (magenta) line
denote the line with $\Delta n_t=r/8$.}\label{figure4}
\end{figure}

Now, let us turn our attention to the constraint of the tensor
spectral index $n_t$. Inserting the best-pivot multipole
$\ell_t^*$ into the second formula in Eq.(15) and taking into
account the corresponding noise power spectra, we obtain the
$\Delta n_t$ as a function of $r$, which are presented in
FIG.\ref{figure4}.  In this figure, we have considered the Planck
and PolarBear (II) experiments. We find that, the value of $\Delta
n_t$ in {\bf Case A} is similar with that in {\bf Case B},
although the Planck noise $N_{\ell}^B$ is nearly 300 times larger
than that of PolarBear (II). When $r>0.08$, Planck can give a
tighter constraint, and when $r<0.08$, PolarBear (II) can give a
tighter constraint. When $r=0.1$, we find $\Delta n_t=0.23$ for
{\bf Case A}, and $\Delta n_t=0.24$ for {\bf Case B}. Both of them
are fairly large for the constraint of the inflationary models.
The single-field slow-roll inflationary models predict the
consistency relation $n_t=-r/8$ \cite{lyth}, which provides the
unique way to exactly test or rule out this kind of models. In
order to answer: whether the observations provide the probability
to check the consistency relation, we can compare the values of
$\Delta n_t$ with $r/8$. If $\Delta n_t<r/8$, we can say the
constraint on $n_t$ is tight enough to check the consistency
relation. From FIG.\ref{figure4}, we find that $\Delta n_t<r/8$ is
satisfied only if $r>0.9$ for {\bf Case A}, and $r>0.8$ for {\bf
Case B}. Unfortunately, these models have been safely excluded by
the current observations \cite{currentconstraint}. So we conclude
that, by either of the single experiment, Planck or PolarBear
(II), we cannot constrain $n_t$ tight enough to check the
consistency relation.


Now, let us consider {\bf Case C}, which has combined the Planck
and PolarBear (II) experiments. We find in this case, the
constraint on $n_t$ becomes much tighter than that in {\bf Case A}
or {\bf B}. Comparing with {\bf Case B}, the value of $\Delta n_t$
is reduced by a factor $2$. When $r=0.1$, $\Delta n_t=0.10$, which
is much smaller than that in {\bf Case A} or {\bf B}. From
FIG.\ref{figure4}, we also find that $\Delta n_t<r/8$ is satisfied
only if $r>0.5$. If considering the contribution of $T,E,C$, i.e.
{\bf Case D}, the constraint on $n_t$ can be  even reduced when
$r>0.2$. So combining the Planck and PolarBear (II) experiments
can effectively reduce the uncertainty of $n_t$, although the
noise power spectra of Planck experiment is much larger than that
in PolarBear (II).


\begin{figure}[t]
\centerline{\includegraphics[width=8cm]{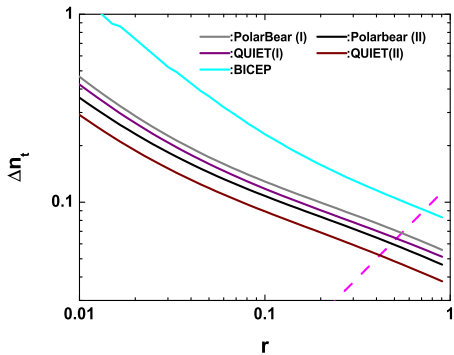}} \caption{The
figure shows the uncertainty $\Delta n_t$ in Case D for the
various ground-based experiments.}\label{figurec3}
\end{figure}

In FIG. \ref{figurec3}, we also plot the $\Delta n_t$ as a function of $r$ in {\bf Case D},
where the various ground-based experiments are considered.  We find that, by combing QUIET (II) and Planck experiments, we can get a
constraint $\Delta n_t=0.09$ for the model with $r=0.1$, and a constraint $\Delta n_t=0.06$ for the model with $r=0.3$.


\begin{figure}[t]
\centerline{\includegraphics[width=9cm]{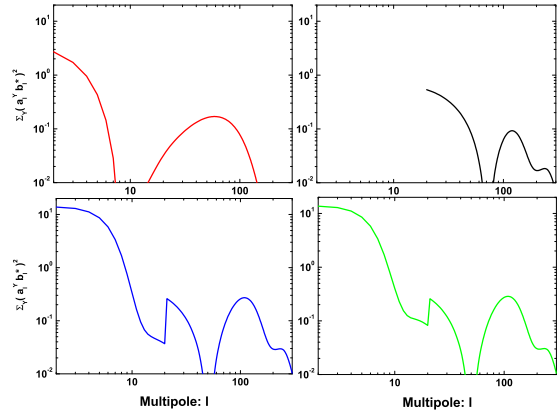}} \caption{In the
model of $r=0.1$ for Planck and PolarBear (II) experiments, we
plot the quantities of $\sum_{Y}(a_{\ell}^Y b_{\ell}^*)^2$ as a
function of multipole $\ell$ in Case A (red line), Case B (black
line), Case C (blue line) and Case D (green line).}\label{figure5}
\end{figure}

Let us investigate the contribution of the observation in the
individual multipole $\ell$. From Eq.(15), we find the quantity
$(1/\Delta n_t)^2$ is a sum of $\sum_Y(a_{\ell}^Yb_{\ell}^*)^2$
for the multipole $\ell$. In FIG.\ref{figure5} we plot the
quantity $\sum_Y(a_{\ell}^Yb_{\ell}^*)^2$ as a function of $\ell$
for all the four cases, where we have considered the model with
parameter $r=0.1$. In all cases, the contribution on the
constraint of $\Delta n_t$ mainly comes from the observation in
the rage $\ell<300$. From this figure, we also find that this
quantity has the zero value when $\ell=\ell_t^*$, due to
$b_\ell^*(\ell=\ell_t^*)=0$. So the observation around the
best-pivot multipole is not important for the constraint of $n_t$.
In the range of $\ell<300$, when $\ell\ll\ell_t^*$ or
$\ell\gg\ell_t^*$, the value of $b_{\ell}^*$ is large, as well as
the quantity $\sum_Y(a_{\ell}^Yb_{\ell}^*)^2$, which gives the
important contribute for the constraint of $n_t$. Especially the
observation in the largest scale. For instance, in {\bf Case C}
the best-pivot multipole $\ell_t^*=51$, so $(b_{\ell=2}^*)^2=10.5$
and $(b_{\ell=200}^*)^2=1.86$, the former one is 6 times larger
than the latter one. So the observation at $\ell=2$ is much more
important than that at $\ell=200$, for constraining the spectral
index $n_t$. We conclude that, the Planck observation in the
largest scale is extremely important for the constraint of $n_t$,
although the noise power spectra of Planck satellite is much
larger than that of the ground-based experiments.



\section{Conclusion\label{s4}}

Detecting the signal of RGWs is one of the most important tasks
for the forthcoming CMB experiments, including the ground-based,
space-based and balloon-borne experiments.  In this letter, by
calculating the uncertainty of the parameters $r$ and $n_t$, we
compared the detection abilities of the upcoming space-based
Planck mission and the various ground-based experiments. Comparing
with Planck experiment, ground-based experiments have the much
smaller instrumental noise power spectra, but can only observe the
CMB field for a small portion of full sky.

We find that ground-based experiments predict a much larger $S/N$
for all the inflationary models, by observing the CMB power
spectra in the intermedial multipole range.  Even we combine it
with Planck experiment, the value of $S/N$ cannot be obviously
improved.

By calculating the value of $\Delta n_t$, we find ground-based
experiments have the similar ability with Planck mission for the
determination of $n_t$.  If combining them, the value of $\Delta
n_t$ can be much reduced, which provides an excellent opportunity
to distinguish the various inflationary models. We also find that,
the observation in the largest scale ($\ell<20$) is extremely
important for determining the spectral index $n_t$.

~

{\bf Acknowledgement:}

W.Zhao thanks NAOC and BNU for invitation and hospitality. This
work is supported by Chinese NSF under grant Nos. 10703005 and
10775119. In this paper, we have used the CAMB code for the
calculation of CMB power spectra \cite{camb}.

\appendix

\section{Instrumental parameters of the various ground-based experiments\label{appendix}}

In this appendix, we shall list the instrumental parameters for
the various ground-based experiment, which including BICEP,
PolarBear (I) and (II), QUIET (I) and (II). The parameters are
detailed listed in Tables \ref{tablebicep}$-$\ref{tablequiet2},
separately.

\begin{table}
\caption{Instrumental parameters for BICEP experiment
\cite{bicep}. }
\begin{center}
\label{tablebicep}
\begin{tabular}{|c|c|c|}
  \hline
   Band center ~[GHz] & 97.7& 151.8  \\
  \hline
  $N_d$  & 50 &   48   \\
  \hline
   NET~[$\mu$K$\cdot$ sec$^{\frac{1}{2}}$] &480 &  420  \\
    \hline
   FWHM~[arcmin]& $55$ &  $37$  \\
       \hline
  \multicolumn{1}{|c|}{$f_{\rm sky}$} & \multicolumn{2}{c|}{$0.024$}\\
         \hline
   \multicolumn{1}{|c|}{$\ell$ range} & \multicolumn{2}{c|}{$20\sim1000$}\\
             \hline
  \multicolumn{1}{|c|}{Integration time} & \multicolumn{2}{c|}{$380$~Days}\\
  \hline
\end{tabular}
\end{center}
\end{table}

\begin{table}
\caption{Instrumental parameters for PolarBear (I) experiment
\cite{polarbear}.  }
\begin{center}
\label{tablepolarbear1}
\begin{tabular}{|c|c|c|c|}
  \hline
   Band center ~[GHz] & 90& 150 & 220 \\
  \hline
  $N_d$  & 104 &   160 & 96  \\
  \hline
   NET~[$\mu$K$\cdot$ sec$^{\frac{1}{2}}$] &220 &  244 & 453 \\
    \hline
   FWHM~[arcmin]& $6.7$ &  $4.0$ & $2.7$ \\
       \hline
  \multicolumn{1}{|c|}{$f_{\rm sky}$} & \multicolumn{3}{c|}{$0.012$}\\
         \hline
   \multicolumn{1}{|c|}{$\ell$ range} & \multicolumn{3}{c|}{$30\sim1000$}\\
             \hline
  \multicolumn{1}{|c|}{Integration time} & \multicolumn{3}{c|}{$0.45$~Years}\\
  \hline
\end{tabular}
\end{center}
\end{table}

\begin{table}
\caption{Instrumental parameters for PolarBear (II) experiment
\cite{polarbear}.  }
\begin{center}
\label{tablepolarbear2}
\begin{tabular}{|c|c|c|c|}
  \hline
   Band center ~[GHz] & 90& 150 & 220 \\
  \hline
  $N_d$  & 400 &   600 & 200  \\
  \hline
   NET~[$\mu$K$\cdot$ sec$^{\frac{1}{2}}$] &220 &  244 & 453 \\
    \hline
   FWHM~[arcmin]& $6.7$ &  $4.0$ & $2.7$ \\
       \hline
  \multicolumn{1}{|c|}{$f_{\rm sky}$} & \multicolumn{3}{c|}{$0.024$}\\
         \hline
   \multicolumn{1}{|c|}{$\ell$ range} & \multicolumn{3}{c|}{$20\sim1000$}\\
             \hline
  \multicolumn{1}{|c|}{Integration time} & \multicolumn{3}{c|}{$0.45$~Years}\\
  \hline
\end{tabular}
\end{center}
\end{table}

\begin{table}
\caption{Instrumental parameters for QUIET (I) experiment
\cite{quiet}.  }
\begin{center}
\label{tablequiet1}
\begin{tabular}{|c|c|c|}
  \hline
   Band center ~[GHz] & 40& 90  \\
  \hline
  $N_d$  & 19 &   91   \\
  \hline
   NET~[$\mu$K$\cdot$ sec$^{\frac{1}{2}}$] &160 &  250  \\
    \hline
   FWHM~[arcmin]& $23$ &  $10$  \\
       \hline
  \multicolumn{1}{|c|}{$f_{\rm sky}$} & \multicolumn{2}{c|}{$0.04$}\\
         \hline
   \multicolumn{1}{|c|}{$\ell$ range} & \multicolumn{2}{c|}{$20\sim1000$}\\
             \hline
  \multicolumn{1}{|c|}{Integration time} & \multicolumn{2}{c|}{$2$~Years}\\
  \hline
\end{tabular}
\end{center}
\end{table}

\begin{table}
\caption{Instrumental parameters for QUIET (II) experiment
\cite{quiet}.  }
\begin{center}
\label{tablequiet2}
\begin{tabular}{|c|c|c|}
  \hline
   Band center ~[GHz] & 40& 90  \\
  \hline
  $N_d$  & 1000 &   1000   \\
  \hline
   NET~[$\mu$K$\cdot$ sec$^{\frac{1}{2}}$] &160 &  250  \\
    \hline
   FWHM~[arcmin]& $23$ &  $10$  \\
       \hline
  \multicolumn{1}{|c|}{$f_{\rm sky}$} & \multicolumn{2}{c|}{$0.04$}\\
         \hline
   \multicolumn{1}{|c|}{$\ell$ range} & \multicolumn{2}{c|}{$20\sim1000$}\\
             \hline
  \multicolumn{1}{|c|}{Integration time} & \multicolumn{2}{c|}{$3$~Years}\\
  \hline
\end{tabular}
\end{center}
\end{table}


\end{document}